# A higher-order implementation of rewriting

Lawrence Paulson

Computer Laboratory
University of Cambridge
Corn Exchange Street
Cambridge CB2 3QG
England

Abstract

Many automatic theorem-provers rely on rewriting. Using theorems as rewrite rules helps to simplify the subgoals that arise during a proof.

LCF is an interactive theorem-prover intended for reasoning about computation. Its implementation of rewriting is presented in detail. LCF provides a family of rewriting functions, and operators to combine them. A succession of functions is described, from pattern matching primitives to the rewriting tool that performs most inferences in LCF proofs.

The design is highly modular. Each function performs a basic, specific task, such as recognizing a certain form of tautology. Each operator implements one method of building a rewriting function from simpler ones. These pieces can be put together in numerous ways, yielding a variety of rewriting strategies.

The approach involves programming with higher-order functions. Rewriting functions are data values, produced by computation on other rewriting functions. The code is in daily use at Cambridge, demonstrating the practical use of functional programming.

## 1. Introduction to rewriting

When trying to prove a theorem, one approach is to simplify it by applying left-to-right *rewrite rules*. For example, consider the proof that addition of natural numbers is associative. We can take the natural numbers to have the form 0, SUCC(0), SUCC(SUCC(0)), ... , and define addition using the axioms

    0 + n = n

    SUCC(m) + n = SUCC(m + n)

Associativity of addition can be stated

    (m+n)+k = m+(n+k)

Induction on m reduces the problem to proving the two goals:

    (0+n)+k = 0+(n+k)                         (base case)

    (SUCC(m)+n) + k = SUCC(m) + (n+k)        (step case)
        with induction hypothesis
        (m+n)+k = m+(n+k)

When conducting such a proof by hand, we would not even write down an expression like 0+(n+k), but would cross out the 0 immediately, proving the base case. For the step case, we would expand out both sides, writing

    left side:
    (SUCC(m)+n) + k =          (by definition of addition)
     SUCC(m+n) + k =           (by definition of addition)
     SUCC((m+n)+k) =           (by induction hypothesis)
     SUCC(m+(n+k))

    right side:
    SUCC(m) + (n+k) =          (by definition of addition)
    SUCC(m+(n+k))

    QED

Note that the cancelling of the 0 in 0+(n+k) can be regarded as the *rewriting* of 0+(n+k), using the first axiom of addition as a rewrite rule. The other addition axiom, and the induction hypothesis, are also used as rewrite rules in the proof.

Rewriting is fundamental to most research on proving theorems by computer [$boyer79,$cohn83,$paulson83a]. Many theorems can be proved by induction and rewriting, or even by rewriting alone. For theories where all axioms are equations, it is sometimes possible to prove additional equations and achieve a *complete set* of rewrite rules, capable of rewriting any term into canonical form [$huet80]. This provides a decision procedure for testing the validity of *any* equation of the theory.

There is a wide variety of rewriting strategies, as Kuechlin [$kuechlin82] thoroughly discusses. In what order should rewrite rules be considered if more than one applies? What about *looping* rewrites such as m+n=n+m ? Should a term be rewritten again and again until no rules apply, or only a bounded number of times? Should a term be traversed top-down or bottom-up, left-to-right or right-to-left? And how can we make use of rewrites that have pre-conditions, such as

if m is not zero then    (m*n)/m = n   ?

This paper presents a family of rewriting primitives, and operators to combine them. It proceeds in stages, from pattern-matching primitives, to instantiation functions, term and formula rewriting functions, tautology solvers, finally discussing the rewriting theorem-prover that is provided in the LCF system [$paulson83a]. The functions at each stage are constructed from those of the previous stage by functional operators that express simple computational intuitions such as sequencing, alternation, and repetition.

Most function definitions are brief and directly express the design decisions such as traversal order. Even the theorem-proving function is only twelve lines, which form a readable summary of its rewriting strategy. If you dislike the standard strategy, you can easily implement another. This modularity is in sharp constrast to large, monolithic implementations of rewriting, such as LCF's previous one.

*Higher-order functions* play a vital role. The rewriting primitives are functions, the operators are functionals. The method is implemented in the programming language ML [$gordon79,$gordon82]. It cannot be implemented in Backus's FP language [$backus78], which allows only first-order functions.

## 2. The LCF proof assistant

LCF is an interactive theorem-prover for Dana Scott's Logic for Computable Functions [$gordon79]. This logic, called PPLAMBDA, provides the "undefined" element UU and the associated partial ordering; it allows reasoning about denotational semantics, programs that manipulate unbounded streams, etc. LCF lets you declare constants and types, assert axioms, and prove theorems; it records all this information in a theory file. You can build a theory on top of other theories, inheriting the constants, axioms, etc. of the other theories. A major verification project might involve a theory hierarchy containing theories of lists, natural numbers, sets, specialized data structures, and theories of increasingly complex functions operating on these types.

LCF lies mid-way between a step-by-step proof checker, and an automatic theorem-prover. When conducting a proof, you are responsible for performing each logical inference. You may automate parts of this task by writing programs in the *meta-language,* ML. The standard theorem-proving tools, such as the rewriting functions, are written in ML. ML treats the terms and formulas of PPLAMBDA as data values, providing functions to build them and take them apart. Theorems are also data values.

PPLAMBDA is a *natural deduction* logic [$manna74]: theorems are proved relative to a set of assumptions. A theorem [$A_1$; ...; $A_n$] |- B means that the *conclusion* B holds whenever the *assumptions* $A_1$, ..., $A_n$ hold. (The conclusion and assumptions are formulas.) In ML, a theorem is a value of the abstract type *thm,* represented by a pair ([$A_1$;...;$A_n$], B). Such a pair is traditionally called a *sequent*.

ML provides functions to decompose a theorem into its conclusion and assumptions, but not to construct an arbitrary theorem. Theorems must be *proved* by applying inference rules to axioms. LCF uses typical introduction/elimination inference rules, implemented as functions whose arguments and results have type thm. Type-checking guarantees that inference rules are only applied to theorems.

Most proofs are conducted *backwards,* starting from the desired goal (expressed as a sequent). The goal is reduced to simpler subgoals, and those reduced to further subgoals, until all the subgoals have been reduced to tautologies. Backwards proof uses *tactics*. A tactic is a function that maps a goal to a list of subgoals, paired with an inference rule. The inference rule justifies the choice of subgoals. Given theorems that assert the subgoals, it produces a theorem that asserts the goal.

Tactics correspond to simple reasoning methods. For instance, the conjunction tactic CONJ_TAC reduces any goal of the form A/\B to the two subgoals A and B; the discharge tactic DISCH_TAC reduces any goal of the form A==>B to the subgoal of proving B under the assumption A. If you apply a tactic to a goal that it cannot handle, then it will *fail* via ML's exception mechanism (described below).

It would be tedious indeed to prove theorems in such tiny steps as CONJ_TAC and DISCH_TAC. Most interactive proofs rely on more powerful tactics, constructed from the primitive ones using functions called *tacticals*. The basic ones are THEN, ORELSE, and REPEAT.

> $tac_1$ THEN $tac_2$
> > applies $tac_1$, then applies $tac_2$ to all resulting subgoals
>
> $tac_1$ ORELSE $tac_2$
> > applies $tac_1$, if it fails then applies $tac_2$
>
> REPEAT tac
> > applies tac recursively on the goal and the resulting subgoals,
> > returning the subgoals for which tac fails

You may be surprised to see the imperative notions of sequencing, alternation, and repetition embodied in higher-order functions that manipulate proof strategies. This paper will show that the same notions apply to rewriting.

## 3. Reference summary of ML

This section describes just enough of ML to enable you to follow the rest of the paper. For instance, ML's polymorphic type system is not discussed because most of the functions in this paper are strongly typed. Gordon [$gordon82] gives a good introduction to ML and its use in theorem-proving. The Edinburgh LCF Manual [$gordon79] also contains an introduction to ML, and is helpful to have at hand if you intend to study the paper in detail.

Note: this paper concerns Cambridge LCF, a revised version of Edinburgh LCF. Most of the changes involve the logic PPLAMBDA [$paulson83c]. Despite incompatibilities, documentation on Edinburgh LCF is still useful.

### 3.1. Values

Values of the language ML include

* the *integers* 0, 1, 2, ..., with operators +, -, *, /, etc.

* the *booleans* true and false, with operators &, or, not, along with the conditional expression:
  if b then x else y

* *tokens*, which are character strings such as 'NO_CONV' or 'Hi Daddy'

* *pairs* such as 0,true -- which may be iterated to form *tuples:* 'manny','moe','jack'

* *lists* [$x_1$; ...; $x_n$], such as [ (0,true) ; (1,false) ]

* *functions,* which may be passed as values, retaining their original variable bindings even outside of the scope where they were created

* *terms, formulas, theorems* of the logic PPLAMBDA

## 3.2. Types

Types include

| | |
|---|---|
| int | (integers) |
| bool | (booleans) |
| tok | (tokens) |
| a # b | (pairs with components of the types a and b) |
| a list | (lists with elements of type a) |
| a -> b | (functions from a to b) |
| term | (PPLAMBDA terms) |
| form | (PPLAMBDA formulas) |
| thm | (PPLAMBDA theorems) |

## 3.3. Declarations

Typed at top level, the declaration

    let x = 3;;

causes x to denote 3 for the rest of the terminal session. Inside a program, the declaration

    let x = 3 in x+x

causes x to denote 3 only in the expression x+x. A local declaration supersedes any global one, using static scope rules.

Declarations include

    let x = 3                      (values)

    let double k = k+k           (functions)

    let times x y = x*y          (curried functions)

    letrec fact n =                (recursive functions)
      if n=0 then 1
      else times n (fact(n-1))

Note that parentheses are not required around arguments to functions. Thus double k means the same as double(k).

### 3.4. Trapping failures

An attempt to perform an illegal operation, such as division by zero, causes ML to signal *failure*. Each failure has an associated error message, or *failure token*. A program can signal failure via the expression

    failwith 'my error'

A failure halts execution unless it is *trapped*. Evaluating the expression x?y computes the value of x; if x fails, then it computes the value of y.

Some people object to failure trapping, comparing it to a goto statement. They prefer to use conditional expressions to prevent failures from occurring. This is impractical for combining proof strategies, where success or failure is difficult to predict. Though implemented as a goto, failure can be understood as an error value that is passed along until it is tested for. A common use of failure is to reject inputs that do not match an expected pattern.

### 3.5. Standard functions and operators

Infix operators include

    . --"conses" an element to the front of a list
        x . $[y_1;...;y_n]$ ---> $[x;y_1;...;y_n]$

    @ --appends two lists
        $[x_1;...;x_m]$ @ $[y_1;...;y_n]$ ---> $[x_1;...;x_m;y_1;...y_n]$

    o --composes two functions
        (f o g)x ---> f(g x)

Functions include

    map  --applies a function to each element of a list
        map f $[x_1;...;x_n]$ ---> $[f\ x_1; ...; f\ x_n]$

    mapfilter  --like map but does not propagate failure; ignores any
        list elements for which f fails

    flat  --flattens a list of lists into a list
        flat $[l_1;...;l_n]$ ---> $l_1$ @ ... @ $l_n$

    itlist  --iterates down a list, accumulating a result
        itlist f $[x_1;...;x_n]$ y ---> f $x_1$ (f ... (f $x_n$ y))

    fst,snd  --select components of pairs
        fst (x,y) ---> x
        snd (x,y) ---> y

## 3.6. Manipulating PPLAMBDA abstract syntax

The terms of PPLAMBDA, which denote computable values, are elements of the ML type *term*. Terms have four abstract syntax classes:

| | |
|---|---|
| C | (constant, where C is a constant symbol) |
| x | (variable) |
| \x.t | (lambda-abstraction) |
| t u | (combination, or function application) |

There is syntactic sugar for certain common expressions:

| | |
|---|---|
| p => t \| u | (conditional expression, actually "COND p t u") |
| t,u | (pair, actually "PAIR t u") |

You may declare constant symbols; the standard ones include

| | |
|---|---|
| UU | (bottom, or undefined element) |
| TT | (true, a computable truth value) |
| FF | (false) |
| FST | (a selector function, like fst in ML) |
| SND | (like snd in ML) |

Note: though not discussed here, PPLAMBDA terms obey a polymorphic type system similar to ML's.

The formulas of PPLAMBDA, which denote logical statements, are elements of the ML type *form*. Formulas have seven abstact syntax classes:

```
!x.A      (universal quantifier)
?x.A      (existential quantifier)
A /\ B    (conjunction)
A \/ B    (disjunction)
A ==> B   (implication)
A <=> B   (if-and-only-if)
P t       (predicate, where P is a predicate symbol)
```

Standard formulas include

```
TRUTH()          (standard tautology predicate)
FALSITY()        (standard contradiction predicate)
t == u    (equivalence of t and u, actually "equiv(t,u)" )
t << u    (Scott inequivalence of t and u, actually "inequiv(t,u)" )
~A        (negation, actually "A ==> FALSITY()")
```

Note: the formula t==u expresses *equivalence* rather than *equality*. Roughly speaking, t and u are equivalent if they are both undefined, or both defined and equal.

The ML expression for constructing a PPLAMBDA object (term or formula) consists of that object enclosed in quotation marks. These are syntax trees, not character strings! ML provides destructor functions for taking apart PPLAMBDA objects.

```
dest_comb  "f x"    --->  "f","x"
dest_equiv "t==u"   --->  "t","u"
dest_conj  "A/\B"   --->  "A","B"
dest_imp   "A==>B"  --->  "A","B"
etc.
```

The theorems of PPLAMBDA, which denote proved formulas, are elements of the ML type *thm*. Inference rules are provided as functions that produce theorems. For instance, the function MP implements Modus Ponens.

```
MP "|-A==>B" "|-A"  --->  "|-B"
```

Note: this example, and others below, show theorems in quotation marks. These depict values rather than expressions. ML does not allow quoted theorems as expressions, since they would construct theorems without proof. Inference rules like MP must be applied to identifiers or other expressions of type thm.

The function *concl* returns the conclusion of a theorem.

```
concl "[A_1;...;A_n] |- B"  --->  "B"
```

## 4. Pattern matching primitives

Now we are ready to examine LCF's rewriting functions, starting with primitives and working upwards. The ML programs below omit most PPLAMBDA inferences, as well as code for optimization and debugging. The most frequently used inferences are "wired in" as additional primitives, to avoid deriving them repeatedly. However, the programs running in LCF are essentially as described here.

A quantified theorem such as |-!x.A stands for an infinity of theorems, one for each x. In a proof, you are likely to need some of these instances, rather than the general form. LCF's pattern matching primitives relieve you of the tedium of instantiating theorems.

### 4.1. Matching terms and formulas

Most rewriting functions depend on the matching functions term_match and form_match. If pattern and object are terms, the call

    term_match pattern object

returns a list of (term,variable) pairs. This expresses the object as an *instance* of the pattern (allowing for renaming of bound variables). If no match exists, term_match fails. The analogous function for formulas is form_match.

Let us look at a terminal session using these functions. The lines beginning with # denote input to LCF, and the other lines denote the response.

We bind a complex term, a conditional, to the ML identifier tm_obj. ML responds by printing the value and its type.

```
#let tm_obj = "TT=> (FF,TT) | (TT,FF)";;
tm_obj = "(TT => (FF,TT) | (TT,FF))" : term
```

A variable can be matched to the term. (The resulting match includes a list for PPLAMBDA types such as ":*", which we ignore for simplicity.)

```
#term_match "x" tm_obj;;
["(TT => (FF,TT) | (TT,FF))","x"],
[":tr # tr",":*"]
: ((term # term) list # (type # type) list)
```

The term can be matched against a conditional, breaking it into parts.

```
#term_match "p=>x|y" tm_obj;;
["TT,FF","y"; "FF,TT","x"; "TT","p"],
[":tr # tr",":*"]
: ((term # term) list # (type # type) list)
```

Since "p=>x|y" is merely syntactic sugar for "COND p x y", we can also match the term against a combination.

```
#term_match "f x" tm_obj;;
["TT,FF","x"; "COND TT(FF,TT)","f"],
[":tr # tr",":**"; ":tr # tr",":*"]
: ((term # term) list # (type # type) list)
```

We cannot match the term against something of a different form.  If we try, term_match fails.

```
#term_match "p=>FF|y" tm_obj;;
evaluation failed     term_match

#term_match "\p.p" tm_obj;;
evaluation failed     term_match
```

We can conduct a similar session using form_match.

```
#let fm_obj = "!x. (x,TT) == UU";;
fm_obj = "!x. (x,TT) == UU" : form

#form_match "!y. (x,y) == UU" fm_obj;;
evaluation failed     form_match

#form_match "?x. (x,TT) == UU" fm_obj;;
evaluation failed     form_match

#form_match "!y. (y,z) == UU" fm_obj;;
["TT","z"],
[":tr",":**"]
: ((term # term) list # (type # type) list)
```

### 4.2.  Instantiating theorems by matching

The functions term_match and form_match are too primitive for most applications.  Their main purpose is to implement the next level of abstraction, which provides functions for instantiating theorems.  Calling

```
PART_TMATCH partfn A t
```

11matches the term t to some part of the theorem A obtained by the function partfn, and returns A with its types and variables instantiated. The partfn is composed from destructor functions such as fst, dest_comb, and dest_equiv. It is applied to the conclusion of the theorem after removing outer universal quantifiers.

PART_TMATCH makes it easy to define inference rules that involve matching. For instance, PPLAMBDA includes an axiom stating that the bottom element is smaller than any other element.

    MINIMAL               |- !x. UU << x

Instantiating MINIMAL is inconvenient (not only x, but PPLAMBDA types must be instantiated), so LCF provides an inference rule MIN that maps any term t to the theorem |-UU<<t. This rule can be expressed using PART_TMATCH and the function (snd o dest_inequiv):

    let MIN = PART_TMATCH (snd o dest_inequiv) MINIMAL;;

Let us see how PART_TMATCH determines what part of MINIMAL to match against. This computation takes place before a term t is applied, since PART_TMATCH is a curried function.

```
(snd o dest_inequiv) (concl "|-UU << x")
snd(dest_inequiv "UU << x")
snd("UU","x")
"x"
```

So MIN matches a term t against the right-hand-side, x, returning the instance |-UU<<t of MINIMAL.

The function PART_FMATCH is analogous, but matches some subformula of the theorem rather than a subterm. One of its applications is a simple resolution rule, MATCH_MP. This is a Modus Ponens that matches an implication to an antecedent.

```
let MATCH_MP impth =
    let match = PART_FMATCH (fst o dest_imp) impth
    in
    \th. MP (match (concl th)) th;;
```

Calling MATCH_MP "|-!$x_1$...$x_n$. A ==> B" "|-A'", where A' is an instance of A, returns the corresponding instance of B. By convention, let us write this instance |-B'. The composite function (fst o dest_imp) takes the first part of an implication, which is the antecedent. One of my proofs [$paulson83a] involves a total function VARS_OF and a theory of strict lists. The function MAP, which maps any function f over a list, produces a total function if f is total. The rule MATCH_MP can prove that the function (MAP VARS_OF) is total.

First we load, from theory files, two theorems about totality.

```
#let VARS_OF_TOTAL = theorem 'VARS_OF' 'VARS_OF_TOTAL';;
VARS_OF_TOTAL = "|-!t. ~ t == UU  ==> ~ VARS_OF t == UU" : thm

#let MAP_TOTAL = theorem 'list_fun' 'MAP_TOTAL';;
MAP_TOTAL =
"|-!f.
   (!x. ~ x == UU  ==> ~ f x == UU) ==>
   (!l. ~ l == UU  ==> ~ MAP f l == UU)"
: thm
```

Using MATCH_MP, we generate more totality theorems.

```
#let TOTAL1 = MATCH_MP MAP_TOTAL VARS_OF_TOTAL;;
TOTAL1 = "|-!l. ~ l == UU  ==> ~ MAP VARS_OF l == UU" : thm

#let TOTAL2 = MATCH_MP MAP_TOTAL TOTAL1;;
TOTAL2 = "|-!l. ~ l == UU  ==> ~ MAP(MAP VARS_OF)l == UU" : thm

#let TOTAL3 = MATCH_MP MAP_TOTAL TOTAL2;;
TOTAL3 = "|-!l. ~ l == UU  ==> ~ MAP(MAP(MAP VARS_OF))l == UU" : thm
```

## 5. Term conversions

The purpose of rewriting is to convert any term t into a term u that is somehow simpler. Furthermore, u must be *proved equivalent* to t. Most theorem-provers take for granted that their rewriting functions are reliable, but the LCF methodology demands that every rewriting step be justified by a theorem.

LCF's rewriting functions are called *conversions*. A term conversion is any function that maps a term t to a theorem |-t==u. This converts the term t to another term u, and proves the two equivalent. Since ML allows us to take theorems apart, we can extract the new term u from the theorem |-t==u.

Let us bind some ML identifiers to terms for use in later terminal sessions. (This simultaneous declaration resembles Lisp's "destructuring let".)

```
#let [abs1; abs2; condu; condt; condf; condfst] = example_terms;;
abs1 = "(\fun.(fun(TT,FF) => x | y))FST" : term
abs2 = "(\t.(\u.t,u)FF)TT" : term
condu = "(UU => (TT,FF,p) | (q,TT,q))" : term
condt = "(TT => x | y)" : term
condf = "(FF => f x | f y)" : term
condfst = "(FST(TT,FF) => x | y)" : term
```

### 5.1. Basic conversions

Beta conversion, BETA_CONV, is standard in LCF. If x is a variable, and u, v are terms, and u[v/x] denotes the substitution of v for x in u, then

BETA_CONV "(\x.u)v"   --->   "|-(\x.u)v == u[v/x]"

This session demonstrates that BETA_CONV performs exactly one beta-conversion, not two or zero.

```
#BETA_CONV abs1;;
"|-(\fun.(fun(TT,FF) => x | y))FST == (FST(TT,FF) => x | y)" : thm

#BETA_CONV abs2;;
"|-(\t.(\u.t,u)FF)TT == (\u.TT,u)FF" : thm

#BETA_CONV condt;;
evaluation failed     BETA_CONV
```

Another basic conversion is to rewrite according to a theorem that states an equivalence. Such theorems are called *rewrites* or *term rewrites*.

REWRITE_CONV "|-!$x_1$...$x_n$. t==u"

This conversion takes any instance of t, such as t', and instantiates the variables $x_1$ ... $x_n$, to return the theorem |-t'==u'. It is implemented using the matching function PART_TMATCH:

```
let REWRITE_CONV = PART_TMATCH (fst o dest_equiv);;
```

Let us explore REWRITE_CONV. First we bind some PPLAMBDA axioms to identifiers, for use in later sessions.

```
#let [COND_UU; COND_TT; COND_FF;
#     MIN_COMB; MIN_ABS;
#     MK_PAIR; FST_PAIR; SND_PAIR]
#  = example_rewrites;;
COND_UU = "|-(UU => x | y) == UU" : thm
COND_TT = "|-(TT => x | y) == x" : thm
COND_FF = "|-(FF => x | y) == y" : thm
MIN_COMB = "|-UU x == UU" : thm
MIN_ABS = "|-\x.UU == UU" : thm
MK_PAIR = "|-FST x,SND x == x" : thm
FST_PAIR = "|-FST(x,y) == x" : thm
SND_PAIR = "|-SND(x,y) == y" : thm
```

We build a conversion from FST_PAIR. It can simplify terms beginning with FST, but not those beginning with SND.

```
#let FST_CONV = REWRITE_CONV FST_PAIR;;
FST_CONV = - : conv

#FST_CONV "FST (TT,FF)";;
"|-FST(TT,FF) == TT" : thm

#FST_CONV "SND (TT,FF)";;
evaluation failed     term_match
```

We build a conversion from COND_TT and use it to simplify a term.

```
#let COND_TT_CONV = REWRITE_CONV COND_TT;;
COND_TT_CONV = - : conv

#COND_TT_CONV condt;;
"|-(TT => x | y) == x" : thm
```

This conversion insists that the condition be TT, not noticing that FST(TT,FF) has the same value.

```
#COND_TT_CONV condfst;;
evaluation failed     term_match
```

## 5.2. Combining conversions

We have operators, similar to tacticals, for combining BETA_CONV and REWRITE_CONV into more powerful conversions.

The operator ORELSEC provides the notion of *alternation*. For conversions $conv_1$ and $conv_2$, and term t, the conversion

$$(conv_1 \text{ ORELSEC } conv_2) \text{ t} \quad \text{--->} \quad conv_1 \text{ t ? } conv_2 \text{ t}$$

It tries $conv_1$; if that fails, then it tries $conv_2$.

Using ORELSEC, we can implement a conversion for conditionals that handles both TT and FF, though still not FST(TT,FF).

```
#let COND_TF_CONV =
#   (REWRITE_CONV COND_TT) ORELSEC
#   (REWRITE_CONV COND_FF);;
COND_TF_CONV = - : conv

#COND_TF_CONV condt;;
"|-(TT => x | y) == x" : thm

#COND_TF_CONV condf;;
"|-(FF => f x | f y) == f y" : thm

#COND_TF_CONV condfst;;
evaluation failed     term_match
```

We can also implement the notion of *sequencing,* defining an operator THENC. For conversions $conv_1$ and $conv_2$, the conversion

    $(conv_1$ THENC $conv_2)$ t

derives

    |-t==t1           (by $conv_1$)
    |-t1==t2  (by $conv_2$)
    and returns
    |-t==t2         (by transitivity, failing if $conv_1$ or $conv_2$ does)

Note that THENC justifies its result using the justifications produced by $conv_1$ and $conv_2$. Using THENC, we can implement a double beta-conversion, which fails if only one beta-conversion is possible.

```
#let BETA_BETA_CONV = BETA_CONV THENC BETA_CONV;;
BETA_BETA_CONV = - : conv

#BETA_BETA_CONV abs1;;
evaluation failed     BETA_CONV

#BETA_BETA_CONV abs2;;
"|-(\t.(\u.t,u)FF)TT == TT,FF" : thm
```

Both ORELSEC and THENC have identity elements. The conversion NO_CONV applies to no terms; it always fails. The conversion ALL_CONV applies to all; it maps any term t to the theorem |-t==t.

```
#NO_CONV  condt;;
evaluation failed    NO_CONV

#ALL_CONV condfst;;
"|-(FST(TT,FF) => x | y) == (FST(TT,FF) => x | y)" : thm
```

For combining several conversions into a multi-way choice, use FIRST_CONV. It is defined using itlist, ORELSEC, and NO_CONV.

>    FIRST_CONV [$conv_1$; ...; $conv_n$]   --->
>      $conv_1$  ORELSEC  ...  ORELSEC  $conv_n$

Using FIRST_CONV, we can implement a conversion for conditionals that handles the conditions UU, FF, and TT.

```
#let COND_CONV =
#   FIRST_CONV (map REWRITE_CONV [COND_TT; COND_FF; COND_UU]);;
COND_CONV = - : conv

#COND_CONV condu;;
"|-(UU => (TT,FF,p) | (q,TT,q)) == UU" : thm

#COND_CONV condf;;
"|-(FF => f x | f y) == f y" : thm

#COND_CONV condt;;
"|-(TT => x | y) == x" : thm

#COND_CONV condfst;;
evaluation failed     FIRST_CONV
```

These operators resemble tacticals. THENC and THEN both express sequencing; ORELSEC and ORELSE both express alternation. The tactic ALL_TAC, which passes on its goal unchanged, is the identity for THEN. (ALL_TAC is called IDTAC in Edinburgh LCF.) The tactic NO_TAC, which fails on all goals, is the identity for ORELSE. ORELSEC and NO_CONV are implemented like ORELSE and NO_TAC, using failure. Most remarkably, we can define *repetition* for conversions exactly as it is defined for tactics.

```
letrec REPEATC conv t =
   ((conv THENC (REPEATC conv))  ORELSEC  ALL_CONV) t;;
```

A fine point: Without the abstraction over t, REPEATC would always loop, because ML uses applicative order (eager) evaluation rather than normal order (lazy) evaluation.

Using REPEATC, we can implement a function that performs as many top-level beta-conversions as possible.

```
#let BETA_N_CONV = REPEATC BETA_CONV;;
BETA_N_CONV = - : conv

#BETA_N_CONV abs1;;
"|-(\fun.(fun(TT,FF) => x | y))FST == (FST(TT,FF) => x | y)" : thm

#BETA_N_CONV abs2;;
"|-(\t.(\u.t,u)FF)TT == TT,FF" : thm

#BETA_N_CONV condt;;
"|-(TT => x | y) == (TT => x | y)" : thm
```

## 5.3. Depth conversions

Now we step beyond the analogy with tacticals, and examine conversions that traverse terms recursively. LCF provides functions for converting subterms: COMB_CONV handles combinations, while ABS_CONV handles abstractions. They fail on terms that do not have the corresponding form.

The conversion (COMB_CONV conv "f t") derives

```
|-f==g            (by conv)
|-t==u            (by conv)
and returns
|-f t == g u      (by substitution)
```

The conversion (ABS_CONV conv "\x.t") derives

```
|-t==u            (by conv)
and returns
|-\x.t == \x.u              (by extensionality, possibly renaming x)
```

Let us combine COMB_CONV and ABS_CONV into a conversion for a term's top-level subterms. Recall that a term can be a constant, variable, abstraction, or combination. Constants and variables are left unchanged, using ALL_CONV:

```
let SUB_CONV conv =
   FIRST_CONV [ COMB_CONV conv;  ABS_CONV conv;  ALL_CONV ];;
```

Now it is simple to write a conversion DEPTH_CONV that recursively rewrites all subterms of a term, in depth-first order.

```
letrec DEPTH_CONV conv t =
   (SUB_CONV (DEPTH_CONV conv) THENC (REPEATC conv)) t;;
```

To try DEPTH_CONV out, we first make a top-level conversion that includes beta-conversion and all our rewrites.

```
#let MANY_CONV =
#   FIRST_CONV (map REWRITE_CONV rewrites)   ORELSEC
#   BETA_CONV;;
MANY_CONV = - : conv
```

Now we make a depth conversion from MANY_CONV and try it on some examples.

```
#let D_CONV = DEPTH_CONV MANY_CONV;;
D_CONV = - : conv

#D_CONV abs2;;
"|-(\t.(\u.t,u)FF)TT == TT,FF" : thm

#D_CONV condfst;;
"|-(FST(TT,FF) => x | y) == x" : thm

#D_CONV abs1;;
"|-(\fun.(fun(TT,FF) => x | y))FST == (FST(TT,FF) => x | y)" : thm
```

We have finally managed to simplify condfst, but what happened with abs1? Clearly its result can be simplified further. We need a more sophisticated conversion, which resimplifies the result of every successful conversion.

```
letrec REDEPTH_CONV conv t =
  (SUB_CONV (REDEPTH_CONV conv) THENC
   ((conv THENC (REDEPTH_CONV conv)) ORELSEC ALL_CONV))
   t;;
```

Here we see that REDEPTH_CONV can simplify abs1 completely.

```
#let RD_CONV = REDEPTH_CONV MANY_CONV;;
RD_CONV = - : conv

#RD_CONV abs1;;
"|-(\fun.(fun(TT,FF) => x | y))FST == x" : thm
```

DEPTH_CONV and REDEPTH_CONV rewrite subterms before rewriting the top-level term. You may prefer TOP_DEPTH_CONV, which tries to rewrite the term before its subterms. This can be quicker, converting FST(x,y) to x without wasting time on y. It can also be slower, converting (\x.F x x)t to (F t t) and then converting t twice.

```
letrec TOP_DEPTH_CONV conv t =
  (REPEATC conv  THENC
   (SUB_CONV (TOP_DEPTH_CONV conv))  THENC
   ((conv THENC (TOP_DEPTH_CONV conv))  ORELSEC ALL_CONV))
  t;;
```

## 6. Interlude

Though we have passed over numerous programs and examples, we are only half-way through Cambridge LCF's implementation of rewriting. Let us pause and reflect on what we have seen so far.

In one sense, there is nothing remarkable about any of the programs above. Pattern matching and rewriting have been around for decades. The old simplifier in Edinburgh LCF [$gordon79] is as powerful as TOP_DEPTH_CONV, and other implementations of rewriting are considerably more elaborate [$kuechlin82].

My approach differs in its modular programming style, which provides both flexibility and readability. The conversions and their operators form a language for expressing rewriting strategies. For instance,

    REPEATC (BETA_CONV THENC BETA_CONV)

evidently performs an even number of beta-conversions. Conversions were developed in order to escape the rigidity of the the old LCF simplifier; now several LCF users are using conversions to suit their particular needs.

Each depth conversion expresses an abstract strategy for the traversal of terms, independent from the conversion to be applied at each subterm. The code for TOP_DEPTH_CONV is a paraphrase of its effect. "Repeatedly apply the conversion conv to the term t as long as possible; then convert recursively the subterms; if the result can still be converted, then convert recursively again."

Conversions may be amenable to algebraic reasoning, since they obey certain identity, associative, and distributive laws. Consider the equations

```
0+a = a+0 = a
(a+b)+c = a+(b+c)
a+b = b+a
a+a = a

1.a = a.1 = a
(a.b).c = a.(b.c)
0.a = a.0 = 0

a.(b+c) = a.b + b.c
(b+c).a = b.a + c.a
```

Researchers [$lehmann77] have begun to study mathematical structures such as *semirings* and *regular algebras*, which satisfy various subsets of these equations. Putting ORELSEC, THENC, NO_CONV,

ALL_CONV for +,.,0,1, it appears that conversions satisfy most of them. The main exception is a+b=b+a; the operator ORELSEC is not symmetric, but always tries its left operand first. Also, the second distributive law fails. The forgoing also applies to tactics.

Most of the research on semirings concerns path-finding in graphs. One can imagine a graph where the nodes are terms and the arcs are conversions. Whether or not this has any practical application, it is important that conversions can be understood through mathematical theories that have arisen in unrelated branches of computer science.

Backus's Turing Award lecture [$backus78] has attracted so much attention that many people now equate functional programming with his FP systems. Indeed, conversions exemplify "changeable parts", "combining forms", and "algebra of programs", which Backus claims as advantages of his technique. However, FP systems allow only first-order programming. FP does not regard functions as data objects; for instance, you cannot build a list of functions. Conversions, and many other parts of LCF, rely heavily on higher-order functions. FP provides only a fixed set of functionals ("combining forms"); a programmer cannot introduce new ones such as TOP_DEPTH_CONV. FP does not allow the trapping of failures, which LCF requires for combining tactics and conversions. It is vital to recognize that the FP style of programming differs fundamentally from the ML style.

## 7. Formula conversions

Now we resume the examination of LCF's rewriting tools. The ideas behind term conversions apply equally well to the rewriting of formulas. Let a *formula conversion* be any function that maps a formula A to a theorem |-A<=>B. This converts A to B and proves the two equivalent. LCF provides a family of formula conversions, and operators to combine them, as for term conversions.

### 7.1. Analogs of term conversions

We can rewrite a formula with a theorem that states a logical equivalence, by invoking the formula conversion

    REWRITE_FCONV "|-!$x_1...x_n$. A <=> B"

Such theorems are called *formula rewrites.* The conversion is implemented like REWRITE_CONV, using the instantiation function PART_FMATCH:

    let REWRITE_FCONV = PART_FMATCH (fst o dest_iff);;

This is useful for expanding out the definition of a predicate, such as

    |- !rel. TRANSITIVE rel  <=>
       !x y z. rel x y == TT  /\  rel y z == TT  ==>  rel x z == TT

LCF provides the identity conversions NO_FCONV, which always fails, and ALL_FCONV, which maps any formula A to |-A<=>A. For sequencing, the conversion (fconv$_1$ THENFC fconv$_2$) is defined in terms of Modus Ponens. The operators ORELSEFC, REPEATFC, and FIRST_FCONV are implemented like their term counterparts.

We can also convert subterms and subformulas. If P is a predicate, then

    PRED_FCONV conv "P(t)"

is a formula conversion that converts the argument t, deriving

    "|-t==u"        (by conv)
    "|-P(t) ==> P(u)"  (by substitution)
    "|-P(u) ==> P(t)"  (by symmetry and substitution)
    and returns
    "|-P(t) <=> P(u)"     (by definition of <=>)

To test PRED_FCONV, we apply it to the depth conversion RD_CONV, from a previous session. The converted formula consists of the predicate P applied to the terms abs2 and condfst.

    #PRED_FCONV RD_CONV "P (^abs2, ^condfst)";;
    "|-P (((\t.(\u.t,u)FF)TT, (FST(TT,FF) => x | y))  <=>  P ((TT,FF),x)" : thm

PPLAMBDA's inference rules allow us to implement conversion operators for the quantifiers and logical connectives. The conversion SUB_FCONV applies a conversion to all top-level terms and formulas of a formula:

```
let SUB_FCONV conv fconv =
  FIRST_FCONV [
    CONJ_FCONV fconv;
      DISJ_FCONV fconv;
      IMP_FCONV fconv;
      IFF_FCONV fconv;
      FORALL_FCONV fconv;
      EXISTS_FCONV fconv;
      PRED_FCONV conv];;
```

For mapping a conversion over all subformulas of a formula, the conversions DEPTH_FCONV, REDEPTH_FCONV, and TOP_DEPTH_FCONV are defined like their term analogs. For example:

```
letrec DEPTH_FCONV conv fconv fm =
  (SUB_FCONV conv (DEPTH_FCONV conv fconv) THENFC
   (REPEATFC fconv))
  fm;;
```

Let us bind some formula rewrites and test formulas for later sessions.

```
#let [P_Q; LESS_UU] = example_frewrites;;
P_Q = "|-!x. P (x,x)  <=>  Q x" : thm
LESS_UU = "|-!x. x << UU  <=>  x == UU" : thm

#let [imp1; conj1; equiv1] = example_forms;;
imp1 =
  "!x y. P ((TT => y | z), SND(y,y))  ==>  Q ((\p.(p => v | y))FF)" : form
conj1 = "?x. x << UU TT  \/  SND(x,TT) == (UU => TT | FF)" : form
equiv1 =
  "!x. ?p. (FST(p,p) => x | UU) == (p => (\z.SND(x,z))x | (\r.r)UU)"
   : form
```

We make a conversion to use our formula rewrites, make a depth conversion from this and the term conversion RD_CONV, and try it on our test data.

```
#let MANY_FCONV =
    FIRST_FCONV (map REWRITE_FCONV [P_Q; LESS_UU]) ;;
MANY_FCONV = - : fconv

#let D_FCONV = DEPTH_FCONV RD_CONV MANY_FCONV;;
D_FCONV = - : fconv

#D_FCONV imp1;;
"|-(!x y. P ((TT => y | z), SND(y,y))  ==>  Q (\p.(p => v | y))FF)  <=>
   (!x y. Q y  ==>  Q y)"
: thm

#D_FCONV conj1;;
"|-(?x. x << UU TT  \/  SND(x,TT) == (UU => TT | FF))  <=>
   (?x. x == UU  \/  TT == UU)"
: thm

#D_FCONV equiv1;;
"|-(!x. ?p. (FST(p,p) => x | UU) == (p => (\z.SND(x,z))x | (\r.r)UU))  <=>
   (!x. ?p. (p => x | UU) == (p => x | UU))"
: thm
```

These formulas are not fully simplified. The next section shows how to eliminate subformulas, such as TT==UU, that are obviously true or false.

## 7.2. Eliminating propositional tautologies

LCF includes conversions that recognize propositional tautologies. For instance, TAUT_CONJ_FCONV can derive

```
TRUTH() /\  A           <=>    A
A /\  TRUTH()           <=>    A
FALSITY() /\ A <=>      FALSITY()
A /\ FALSITY() <=>      FALSITY()
```

Most of the tautology conversion functions are hand-coded to treat a particular class of formulas. But ones for quantifiers are implemented in terms of formula conversions. LCF has stored the theorems

```
FORALL_TRUTH    |- (!x.TRUTH())  <=>  TRUTH()

FORALL_FALSITY  |- (!x.FALSITY()) <=> FALSITY()
```

Using these, the "forall" tautology conversion can simplify !x.TRUTH() and !x.FALSITY(). Its ML definition is

```
let TAUT_FORALL_FCONV =
   (REWRITE_FCONV FORALL_TRUTH)
     ORELSEFC
   (REWRITE_FCONV FORALL_FALSITY);;
```

The family of conversions is modular. To improve the tautology test, write a better version of TAUT_FORALL_FCONV. Perhaps it should simplify !x.A to A for any formula A that does not contain x.

LCF provides the conversion BASIC_TAUT_FCONV, which tries all the tautology tests in turn, failing if none apply.

```
let BASIC_TAUT_FCONV =
  FIRST_FCONV [
       TAUT_CONJ_FCONV;
       TAUT_DISJ_FCONV;
       TAUT_IMP_FCONV;
       TAUT_IFF_FCONV;
       TAUT_FORALL_FCONV;
       TAUT_EXISTS_FCONV;
       TAUT_PRED_FCONV];;
```

There are many ways of building simplifiers from these conversions. The standard one, BASIC_FCONV, uses TOP_DEPTH_CONV to simplify terms, TOP_DEPTH_FCONV to simplify formulas, and BASIC_TAUT_FCONV to find tautologies in the resulting formulas. Many factors play a role in tailoring a simplifier to a specific problem. For instance, the REDEPTH conversions are slower but more thorough than the DEPTH ones. For current LCF applications, the TOP_DEPTH conversions seem to offer the best compromise of speed and generality.

```
let BASIC_FCONV conv fconv =
   TOP_DEPTH_FCONV (TOP_DEPTH_CONV conv) (fconv ORELSEFC BASIC_TAUT_FCONV);;
```

The following session shows how BASIC_FCONV combines our top-level conversions, MANY_CONV and MANY_FCONV. The resulting formula conversion solves the tautologies that were missed before.

```
#let B_FCONV = BASIC_FCONV MANY_CONV MANY_FCONV;;
B_FCONV = - : fconv

#B_FCONV imp1;;
"|-(!x y. P ((TT => y | z), SND(y,y))  ==>  Q (\p.(p => v | y))FF)  <=>
  TRUTH ()"
: thm

#B_FCONV conj1;;
"|-(?x. x << UU TT  \/  SND(x,TT) == (UU => TT | FF))  <=>
  (?x. x == UU)"
: thm

#B_FCONV equiv1;;
"|-(!x. ?p. (FST(p,p) => x | UU) == (p => (\z.SND(x,z))x | (\r.r)UU))  <=>
  TRUTH ()"
: thm
```

## 8. Anatomy of a rewriting tactic

In studying these rewriting functions, let us remain aware of their original purpose: to prove theorems. Now we will study the LCF tactic REWRITE_TAC, which simplifies a goal by rewriting it and removing tautologies. This tactic is evolving over time. Though the version described here is not the latest, its structure illustrates the practical use of conversions and higher-order functions.

### 8.1. Implicative rewrites

The functions REWRITE_CONV and REWRITE_FCONV accept rewriting theorems of the form |-t==u or |-A<=>B. However, there are many equivalences that only hold under certain conditions. Consider a theory of lists with strict CONS and MAP functions. The following theorem holds only by virtue of the antecedent forcing x to be defined in the equivalence:

```
~x==UU  ==>  MAP f (CONS x l) == CONS (f x) (MAP f l)
```

If x is UU, l is NIL, and f is the constant function \y.TT, then the equivalence does not hold:

```
MAP (\y.TT) (CONS UU NIL) == CONS ((\y.TT) UU) (MAP (\y.TT) NIL)
<=>                              (by strictness of CONS, definition of MAP)
MAP (\y.TT) UU == CONS ((\y.TT) UU) NIL
<=>                              (by beta-conversion, strictness of MAP)
UU == CONS TT NIL
<=>                              (by totality of CONS)
FALSITY()
```

In general, these *implicative rewrites* may depend on more than one antecedent. LCF presumes them to have the form, for non-negative n,

$$A_1 \Longrightarrow (... (A_n \Longrightarrow t \equiv u) ...)$$

$$A_1 \Longrightarrow (... (A_n \Longrightarrow (B \Longleftrightarrow C)) ...)$$

How can a conversion use such a theorem, given an instance t' of the left hand term t? If it can prove the instances of the antecedents, $A'_1, ..., A'_n$, then, by Modus Ponens, it can return the theorem |-t'≡u'. How should it try to prove the antecedents? The simplifiers in both Edinburgh LCF and the Boyer/Moore Theorem Prover [$boyer79] solve antecedents by recursively invoking the simplifier. However, there is no need to commit ourselves; we can pass any proof tactic as an argument. Conversions that attempt to prove instances of the antecedents using a tactic tac are

IMP_REW_CONV tac  "|-$A_1 \Longrightarrow$ (... ($A_n \Longrightarrow$ t≡u) ...)"

IMP_REW_FCONV tac  "|-$A_1 \Longrightarrow$ (... ($A_n \Longrightarrow$ (B<=>C)) ...)"

## 8.2. Backwards chaining

Although the latest version of REWRITE_TAC invokes itself to prove the antecedents of implicative rewrites, we will examine an earlier, simpler version. It uses *backwards chaining* -- a proof search that resembles the execution of PPLAMBDA implications as a PROLOG program [$clocksin81]. It is implemented using PART_FMATCH to match the consequent of an implication.

Let us see how backwards chaining can solve antecedents of implicative rewrites. Typically, an antecedent will require that a list l be defined: ˜l≡UU. Consider a first-order theory of lists, with NIL, a strict CONS, and an infix operator APP to append lists. The theory includes theorems asserting that these constants create defined lists:

```
NIL_DEFINED   |- ˜ NIL == UU
CONS_DEFINED  |- ˜ a==UU ==> (˜ l==UU ==> ˜ CONS a l == UU)
APP_DEFINED   |- ˜ l₁==UU ==> (˜ l₂==UU ==> ˜ l₁ APP l₂ == UU)
```

The tactic IMP_SEARCH_TAC performs a depth-first search using a list of such theorems. It searches the list for a theorem whose consequent matches the goal, failing if there is none. Suppose there is a theorem

|-$A_1$==>...==>$A_n$==>B,

and that the goal is an instance B' of B. The tactic calls itself recursively to prove the instances of the antecedents, $A'_1...A'_n$, and proves the goal B' by Modus Ponens. For the search to terminate successfully, there must be theorems with no antecedents (n=0). A theorem such as |-A==>A will cause infinite regress.

In this example, it establishes that a complex list is defined, because it is constructed from the constants NIL, CONS, APP. The variables t, u, and l are all assumed to be defined. The goal tree shows how the initial goal is reduced to simpler subgoals, until the leaves are all true:

~(CONS t l) APP (CONS t (CONS u NIL))) == UU

~CONS t l == UU          ~CONS t (CONS u NIL) == UU

~t==UU      ~l==UU    ~t==UU            ~CONS u NIL == UU

~u==UU    ~NIL==UU

### 8.3. Canonical forms

A predicate logic such as PPLAMBDA allows many different ways of saying the same thing. For instance, (A/\B)==>C is logically equivalent to A==>(B==>C), though IMP_REW_CONV and IMP_SEARCH_TAC expect the latter. Edinburgh LCF addresses this problem by forcing *every* formula into a standard canonical form. Cambridge LCF, described here, takes a more flexible approach. It provides functions for putting theorems into canonical form; the user may invoke these or implement different ones in ML. These functions are *inference rules*. They do not simply manipulate data structures, but prove their output theorems from their input theorems.

The function IMP_CANON converts a theorem into a list of implications. This form is useful in many LCF situations. (Note: bound variables may be renamed; assumptions of the input theorem are passed to the output.)

```
IMP_CANON "|-A /\ B"       --->   (IMP_CANON "|-A") @ (IMP_CANON "|-B")
IMP_CANON "|-(?x.A) ==>B"  --->   IMP_CANON "|-A[y/x] ==> B"
IMP_CANON "|-!x.A"         --->   IMP_CANON "|-A[y/x]"
IMP_CANON "|-(A/\B) ==>C"  --->   IMP_CANON "|-A==>(B==>C)"
IMP_CANON "|-(A\/B) ==>C"  --->   (IMP_CANON "|-A==>C") @ (IMP_CANON "|-B==>C")
```

If the cases above do not apply, and the argument is an implication |-A==>B, then IMP_CANON calls itself on the equivalent theorem, A|-B. This breaks B into a list [A|-B$_1$; ...; A|-B$_n$] of theorems that assume A. It discharges A from each of these, to return [A==>B$_1$; ...; A==>B$_n$]. A theorem that satisfies no cases, such as |-A\/B, passes through unchanged.

In this session, we see IMP_CANON converting several slightly different theorems into the same one.

```
#IMP_CANON (ASSUME "!x. ˜ x==UU  ==>  !y. ˜ y==UU  ==> ˜ f x y ==UU");;
[.|-"˜ x == UU  ==>  ˜ y == UU  ==>  ˜ f x y == UU"] : thm list

#IMP_CANON (ASSUME "!x y. ˜ x==UU  ==> ˜ y==UU  ==> ˜ f x y ==UU");;
[.|-"˜ x == UU  ==>  ˜ y == UU  ==>  ˜ f x y == UU"] : thm list

#IMP_CANON (ASSUME "(˜ x==UU  /\ ˜ y==UU)  ==> ˜ f x y ==UU");;
[.|-"˜ x == UU  ==>  ˜ y == UU  ==>  ˜ f x y == UU"] : thm list

#IMP_CANON (ASSUME "(˜ x==UU  /\ ˜ y==UU)  ==>
#                   !f g. ˜ f x y ==UU /\  ˜ g x y == UU");;
[.|-"˜ x == UU  ==>  ˜ y == UU  ==>  ˜ f x y == UU";
 .|-"˜ x == UU  ==>  ˜ y == UU  ==>  ˜ g x y == UU"]
: thm list
```

Such implications are fine for IMP_REW_CONV and IMP_SEARCH_TAC, but IMP_REW_FCONV expects theorems of the uncommon form C<=>D. The inference rule FCONV_CANON alters the consequent of certain implications into logical equivalences:

```
P(x)      --->    P(x) <=> TRUTH()       (P a predicate symbol)
˜ P(x)    --->    P(x) <=> FALSITY()
C<=>D     --->    unchanged
else fail
```

Thus FCONV_CANON proves logical equivalences that rewrite predicates to TRUTH() and negated predicates to FALSITY(). It passes on any formula rewrites it encounters. For instance, the theorems NIL_DEFINED, CONS_DEFINED, and APP_DEFINED become implicative formula rewrites. (Recall that the formula t==u is shorthand for the predicate equiv(t,u).)

```
|- NIL == UU  <=>  FALSITY
|- ˜ a==UU  ==>  (˜ l==UU  ==>  (CONS a l == UU  <=>  FALSITY()))
|- ˜ l₁==UU  ==>  (˜ l₂==UU  ==>  (l₁ APP l₂ == UU  <=>  FALSITY()))
```

These theorems can solve the antecedents of implicative rewrites using rewriting instead of backwards chaining. The effect is similar to the derivation tree shown at the end of the previous section. TOP_DEPTH_FCONV tries these formula rewrites *before* descending into the antecedent to apply term rewrites.

Thus, if the antecedent can be solved by backwards chaining, it can also be solved by rewriting, with comparable efficiency. Rewriting can solve many more antecedents than backwards chaining can. Its main drawback is a greater danger of infinite regress, trying to rewrite the antecedent of the antecedent of the ... . Any implicative rewrite |-A(t) ==> t==u can cause such regress, invoking itself in the attempt to prove its own antecedent. Again we see that many considerations guide the selection of components when building a simplifier.

### 8.4. The primitive conversion tactic

Though there are many ways to build a conversion, there is only one obvious way to reduce a goal using a conversion. The tactic (FCONV_TAC fconv) uses fconv to convert a goal A to a subgoal B, leaving its assumptions unchanged. If B is just TRUTH(), then FCONV_TAC has achieved the goal A, and returns an empty subgoal list.

### 8.5. The rewriting tactic

The tactic REWRITE_TAC requires all the above pieces. It accepts a list of theorems, and puts them into canonical form using IMP_CANON and FCONV_CANON. It handles implicative rewrites using IMP_REW_CONV and IMP_REW_FCONV, which fail on unacceptable theorems. The function mapfilter gathers the successful conversions; these, along with BETA_CONV, are combined using FIRST_CONV, FIRST_FCONV, and BASIC_FCONV. REWRITE_TAC solves antecedents of implicative rewrites by backwards chaining, using the tactic IMP_SEARCH_TAC. To solve trivial subgoals in chaining, it augments the list of theorems with LCF's reflexivity axiom:

```
EQ_REFL              |- !x.x==x
```

The tactic ASM_REWRITE_TAC calls REWRITE_TAC, using the tactical ASSUM_LIST to append the goal's assumptions to the input list of theorems. Most proofs rely on ASM_REWRITE_TAC for rewriting, in order to take advantage of the assumptions.

```
let REWRITE_TAC thl =
   let thms = flat (map IMP_CANON thl) in
   let chain_tac = IMP_SEARCH_TAC (EQ_REFL . thms) in
   let conv =
         FIRST_CONV (mapfilter (IMP_REW_CONV chain_tac) thms)
         ORELSEC  BETA_CONV
   in
   let fconv =
         FIRST_FCONV
            (mapfilter ((IMP_REW_FCONV chain_tac) o FCONV_CANON) thms)
   in
   FCONV_TAC (BASIC_FCONV conv fconv);;

let ASM_REWRITE_TAC thl =
   ASSUM_LIST (\asl. REWRITE_TAC (asl @ thl));;
```

## 9. Examples of solving goals by rewriting

To see ASM_REWRITE_TAC in use, consider a recent proof of mine [$paulson83a]. It uses a data structure for expressions composed of constants, variables, and combinations of other expressions. It concerns infix functions OCCS and OCCS_EQ; these are relations because they return a truth-valued result. The relation "t OCCS u" searches u for an occurrence of t, returning TT if it finds one. The relation

OCCS_EQ is the reflexive closure of OCCS. They are defined in terms of a boolean operator OR, and a computable equality relation =.

```
let OCCS_EQ =
new_axiom ('OCCS_EQ',
   "!t u. t OCCS_EQ u  ==  (t=u) OR (t OCCS u)");;

let OCCS_CLAUSES =
new_axiom ('OCCS_CLAUSES',
"!t.  t OCCS UU == UU
  /\
  (!c. ~c==UU  ==>
    t OCCS (CONST c) == FF)
  /\
  (!v. ~v==UU  ==>
    t OCCS (VAR v) == FF)
  /\
  (!t₁ t₂. ~t₁==UU  ==>  ~t₂==UU  ==>
    t OCCS (COMB t₁ t₂) == (t OCCS_EQ t₁) OR (t OCCS_EQ t₂))");;
```

We will see how ASM_REWRITE_TAC helps to prove that the relation OCCS is transitive:

```
!t. ~t==UU  ==>
   !u. t OCCS u == TT  ==>
       !u'. u OCCS u' == TT  ==> t OCCS u' == TT
```

Inducting on the variable u' yields four subgoals.

```
#expand (TERM_TAC "u'");;
OK..
4 subgoals
"u OCCS (COMB t₁ t₂) == TT  ==>  t OCCS (COMB t₁ t₂) == TT"
    [ "~ t == UU" ]
    [ "t OCCS u == TT" ]
    [ "u OCCS t₁ == TT  ==>  t OCCS t₁ == TT" ]
    [ "u OCCS t₂ == TT  ==>  t OCCS t₂ == TT" ]
    [ "~ t₁ == UU" ]
    [ "~ t₂ == UU" ]

"u OCCS (VAR v) == TT  ==>  t OCCS (VAR v) == TT"
    [ "~ t == UU" ]
    [ "t OCCS u == TT" ]
    [ "~ v == UU" ]
```

```
"u OCCS (CONST c) == TT  ==>  t OCCS (CONST c) == TT"
   [ "~ t == UU" ]
   [ "t OCCS u == TT" ]
   [ "~ c == UU" ]

"u OCCS UU == TT  ==>  t OCCS UU == TT"
   [ "~ t == UU" ]
   [ "t OCCS u == TT" ]
```

Compare these with the axiom OCCS_CLAUSES; three of them contradict the antecedent, u OCCS u' == TT. Using the axioms OCCS_CLAUSES and OCCS_EQ, the tactic ASM_REWRITE_TAC solves the three easy goals. During initialization, it splits OCCS_CLAUSES into UU, CONST, VAR, and COMB clauses, each an implicative rewrite. While rewriting the goal involving CONST, it notices the assumption ~c==UU, and rewrites u OCCS (CONST c) to FF. Then it rewrites the antecedent FF==TT to FALSITY(). Similarly, it rewrites the consequent to FALSITY(), yielding the trivial goal FALSITY() ==> FALSITY(). ASM_REWRITE_TAC solves the goals involving UU and VAR in the same way. The fourth goal is difficult, but the tactic advances it considerably:

```
"((u = t₁) OR (u OCCS t₁)) OR
 ((u = t₂) OR (u OCCS t₂))     == TT
 ==>
 ((t = t₁) OR (t OCCS t₁)) OR
 ((t = t₂) OR (t OCCS t₂))     == TT"
   [ "~ t == UU" ]
   [ "t OCCS u == TT" ]
   [ "u OCCS t₁ == TT  ==>  t OCCS t₁ == TT" ]
   [ "u OCCS t₂ == TT  ==>  t OCCS t₂ == TT" ]
   [ "~ t₁ == UU" ]
   [ "~ t₂ == UU" ]
```

My paper [$paulson83a] describes the rest of the proof, involving a case split followed by a further call to ASM_REWRITE_TAC.

### 9.1. Recent improvements to REWRITE_TAC

REWRITE_TAC has evolved since this paper was first written. The version presented above is simpler than the latest one. Now we examine, in less detail, the recent improvements. The first implementation contained lots of messy code; I later realized how to express this in a modular style, using conversions.

The main innovation is to use *local assumptions* during rewriting. In the formulas A/\B and A==>B, it is legitimate to assume A when rewriting B. Informally, the truth value of B is irrelevant unless A holds. A more convincing justification is the following derived inference rule, implemented in ML using the primitive rules:

$$\frac{[A;\ A_2] \quad \begin{array}{c} A <=> A_2 \\ B <=> B_2 \end{array}}{\begin{array}{c} (A \wedge B) <=> (A_2 \wedge B_2) \\ (A ==> B) <=> (A_2 ==> B_2) \end{array}}$$

This rule allows us to convert the formula A to $A_2$, then assume these while converting B to $B_2$, without these assumptions appearing in the resulting theorems.

Local assumptions facilitate many proofs. The older rewriting tactic could return a goal containing the formula

    p==TT ==> p=>x|y == x

Rewriting the conditional with the local assumption p==TT simplifies the formula to TRUTH(). If the rewriting tactic cannot make local assumptions, then the user must manipulate the antecedent p==TT into a global position. These manipulations obscure the proof. They reflect the context of the formula, yet the formula is true in any context.

It is not obvious how the conversion (IMP_FCONV fconv) can make local assumptions. Somehow the assumption p==TT must be incorporated into fconv, itself built from other operators. My solution is to supply an additional argument, fconv_fun, which maps a formula to a conversion. The local version of IMP_FCONV is

    LOCAL_IMP_FCONV fconv fconv_fun

When converting an implication A==>B, it uses fconv to convert A to $A_2$, applies fconv_fun to $A_2$ to produce a new conversion, and converts B using that. Likewise there is a LOCAL_CONJ_FCONV. Combining these conversions with those for the other connectives yields a LOCAL_SUB_FCONV, a LOCAL_TOP_DEPTH_FCONV, and a LOCAL_BASIC_FCONV. These local conversions resemble their predecessors, but take the additional argument fconv_fun.

The new REWRITE_TAC is implemented in terms of a highly recursive formula conversion. For solving implicative rewrites, this conversion uses a call to itself rather than IMP_SEARCH_TAC. It passes another recursive call as the fconv_fun argument of LOCAL_SUB_FCONV. For fast pattern matching, it stores rewrites in *discrimination nets* [$charniak80] instead of lists.

A separate improvement is to expand out disjunctions during rewriting, causing automatic case splits. The conversion EXPAND_DISJ_FCONV derives

    (A\/B) ==> C    <=>    (A==>C) /\ (B==>C)
    (A\/B) /\ C    <=>    (A/\C) \/ (B/\C)
    C /\ (A\/B)    <=>    (C/\A) \/ (C/\B)

This is applied at the same point as the tautology conversions. Existential quantifiers are similarly expanded. Expansion of disjunctions is especially effective when A and B, as local assumptions, help to rewrite C.

## 10. Conclusions

You may be thinking, "Conversions seem interesting, but must be hopelessly inefficient." Conversions are heavily used in Cambridge LCF, where they are efficient enough to prove difficult theorems. Runtime ranges from ten seconds to several minutes on a VAX 750 computer.

It is hard to improve the efficiency in the LCF framework. Any simplifier must produce a theorem to justify its result. It must coexist with other theorem-proving tools, and with ML. This precludes some optimizations, such as reducing the number of substitutions by maintaining a global environment of variable bindings [$boyer79]. The current ML compiler generates poor code, though efficient implementations are being developed [$cardelli83].

Conversion functions have many advantages over Edinburgh LCF's simplifier, a large and inscrutable ML program. The operators PART_FMATCH, REWRITE_CONV, TAUT_CONJ_CONV, IMP_CANON, etc., carry out small, well-defined tasks. They have simple specifications and implementations. Together they express the rewriting tactic, REWRITE_TAC, in only a dozen lines.

REWRITE_TAC performs the vast majority of inferences in LCF proofs. Its limitations and abilities are easy to grasp, thanks to its modular structure. This helps the user to plan interactive sessions, and to read tactical proofs like a summary of the hundreds of formal inferences.

Conversions illustrate the power of higher-order functions. Because ML treats functions as first-class data, we can implement rewriting tools as functions and write operators to combine them. Proof tactics are functions too; the conversion IMP_REW_CONV generates and proves subgoals using a tactic passed to it as an argument. The instantiation function PART_TMATCH accepts a function argument that tells it what part of a theorem to match. This programming style differs greatly from the style that Backus [$backus78] recommends, where only first-order functions are allowed.

LCF is used not only for performing particular proofs, but also for research into proof techniques. The programming language ML provides the flexibility needed for this research. The discovery of the operators THEN, ORELSE, and REPEAT, for combining tactics, was a breakthrough in the development of Edinburgh LCF. So it is exciting to find similar operators for combining conversions, especially since tactics and conversions have little else in common. This calls for an inquiry into other instances of this programming style.

*Acknowlegements.* I would like to thank J. Fairbairn, M. Gordon, D. Matthews, and the referees for their detailed comments on the paper. D. Benson introduced me to semirings. This research is supported by the Science Research Council of Great Britain.